\documentclass[10pt,tightenlines,eqsecnum,floats,aps,amsmath,
amssymb,nofootinbib,prd,noshowpacs,notitlepage]{revtex4-1}
\usepackage[utf8]{inputenc}
\usepackage{amsmath,amsthm,latexsym,amssymb,amsfonts,color,graphicx}
\usepackage{hyperref}

\begin{document}

\title{Discretization independence implies non--locality in 4D discrete quantum gravity}

\author{Bianca Dittrich${}^{1}$}
\email{bdittrich@perimeterinstitute.ca}
\author{Wojciech Kami\'nski${}^{1,2}$}
\email{wkaminsk@fuw.edu.pl}
\author{Sebastian Steinhaus${}^{1}$}
\email{steinhaus.sebastian@gmail.com}

\affiliation{
  ${}^{1}$Perimeter Institute for Theoretical Physics, 31 Caroline Street North
  Waterloo, Ontario Canada N2L 2Y5\\
  ${}^{2}$Instytut Fizyki Teoretycznej, Uniwersytet Warszawski,
  ul. Ho\.{z}a 69, 00-681 Warszawa, Poland
  }

\begin{abstract}
The 4D Regge action is invariant under 5--1 and 4--2 Pachner moves, which define a subset of (local) changes of the triangulation. Given this fact one might hope to find a local path integral measure that makes the quantum theory invariant under these moves and hence makes the theory partially triangulation invariant. We show that such a local invariant path integral measure does not exist for the 4D linearized Regge theory. 

To this end we uncover an interesting geometric interpretation for the Hessian of the 4D Regge action. This geometric interpretation will allow us to prove that the determinant of the Hessian of the 4D Regge action does not factorize over $4$--simplices or subsimplices. It furthermore allows to determine configurations where this Hessian vanishes, which only appears to be the case in degenerate backgrounds or if one allows for different orientations of the simplices.

We suggest a non--local measure factor that absorbs the non--local part of the determinant of the Hessian under 5--1 moves as well as a local measure factor that is preserved for  very special  configurations.

\end{abstract}

\maketitle

\section{Introduction}

Many quantum gravity approaches rely on a path integral construction as their foundation, for example spin foam models \cite{spinfoams}, group and tensorial field theory \cite{gft}, (causal) dynamical triangulations \cite{cdt} or quantum Regge calculus \cite{regge}. These approaches have the same goal, namely to provide a way to compute (and give meaning to) the gravitational path integral, i.e. the sum over all histories between two 3--dimensional boundary geometries, where each history is a 4--geometry describing a possible transition weighted by the (exponential of the) Einstein--Hilbert action. An essential part in this path integral is the measure over the space of geometries, i.e. the space of all metrics modulo diffeomorphisms.

To propose a well--defined path integral, one generically has to introduce a regulator to truncate the degrees of freedom of the theory. In gravitational theories mentioned above this is achieved by discretizing the theory, e.g. on a triangulation. However, the introduction of discretizations comes with a caveat: In general, a discretization of the classical theory cannot be chosen uniquely, if the only requirement is that this discretization leads to the correct continuum action.
Whereas some agreement has been reached on the Regge action as a preferred discrete action \cite{schrader}, at least for the theory without cosmological constant, the debate on the measure in Regge calculus \cite{hamber,hamber2,menotti,regge-lund} and spin foams \cite{bojowald,kkl} is not settled.

Even more troubling in the context of gravity, discretizations generically break diffeomorphism symmetry \cite{diffeo-review,broken-symmetry}, which is deeply intertwined with the dynamics of general relativity. Furthermore, this might induce an unphysical dependency of this theory on the choice of the discretization. Different approaches to quantum gravity differ in how they deal with these problems, e.g. in Regge calculus one considers only one triangulation with varying edge lengths, whereas in causal dynamical triangulations one keeps equilateral simplices and sums over all triangulations. Group field theories additionally sum over all topologies. Which of these schemes leads to a sensible theory of quantum gravity cannot be determined a priori.

These intricacies are deeply rooted in the fact that diffeomorphism symmetry is broken by the discretisation and that the relation between discrete and continuous gravity is still hardly understood. This particularly affects the choice of measure in quantum gravity theories, which is crucially important for the dynamics in the continuum limit, since it also resembles a choice of the measure on the space of geometries. 

For spin foam models, arguments that link diffeomorphism symmetry, choice of (anomaly free) measure and divergence structure due to having non--compact gauge orbits from the diffeomorphism group have been made in \cite{louapre,bojowald}.  This led also to the suggestion to choose a measure which has a certain (weak) notion of discretization independence, e.g. such that the amplitudes become independent under `trivial' edge and face subdivisions \cite{bahr,warsaw,kkl,bonzomdittrich}. As mentioned the choice of measure heavily influences the divergence structure of the models \cite{perini,aldo,bonzomdittrich}. Thus one can adjust the measure to obtain the divergence structure that fits the divergences one expects from diffeomorphism symmetry. This of course  does not fully guarantee diffeomorphism symmetry or triangulation independence, as divergences might also arise due to other reasons.

In Regge calculus, several measures have been proposed: in \cite{hamber} Hamber and Williams propose a discretization of the formal continuum path integral, with a local discretization of the (DeWitt) measure \cite{hamber,hamber2}, conflicting with the proposal by Menotti and Peirano \cite{menotti}, who 
mod out a subgroup of the continuum diffeomorphisms resulting in a highly non--local measure. A different discretization, also leading to a non--local measure due to discretizing first the DeWitt super metric \cite{deWitt} and then forming the determinant, was proposed in  \cite{regge-lund}.  

In this work we will pick up the suggestion in \cite{steinhaus11}, to choose a measure that, at least for the linearized (Regge) theory, leads to as much discretization invariance as possible. These considerations require to actually integrate out degrees of freedom and thus take the dynamics into account. 

The requirement of discretisation independence seems to be at odds with interacting theories, which possess local / propagating degrees of freedom. This apparent contradiction can be resolved by allowing a non--local action or non--local amplitudes for the quantum theory -- which in fact are unavoidable if one coarse grains the theory. Non--local amplitudes are however difficult to deal with. We therefore ask in this paper the question, whether in the quantum theory we can retain as much symmetry as in the classical theory, with a choice of local measure. The classical 4D Regge action is known to be invariant under $5-1$ moves and $4-2$ moves, but not under $3-3$ moves \cite{steinhaus11}. Here the non--invariance under the $3-3$ moves -- in fact the only move involving bulk curvature for the solution -- allows the local Regge action to nevertheless lead to a theory with propagating degrees of freedom. We therefore ask whether it is possible to have a local measure for linearized Regge calculus that leads to invariance under $5-1$ and / or $4-2$ moves. Such a measure would therefore reproduce the symmetry properties of the Regge action. We will however show that such a local path integral measure does not exist.


Our requirement of (maximal) discretisation independence is motivated by the `perfect action / discretisation' approach \cite{perfect-action} that targets to construct a discretisation, which `perfectly' encodes the continuum dynamics and has a discrete remnant of the continuum diffeomorphism symmetry. Examples of such `perfect discretisations' are 3D Regge calculus with and without a cosmological constant \cite{Regge-cosmo} and also 4D Regge calculus, if the boundary data impose a flat solution in the bulk. In these examples, the basic building blocks mimic the continuum dynamics, e.g. one takes constantly curved tetrahedra for 3D gravity with a cosmological constant. Such perfect discretisations can be constructed as the fixed point of a coarse graining scheme, 
see for instance \cite{marsden,Regge-cosmo,anharmonic}.

Once such a discretisation is constructed, the predictions of the theory become independent of the fineness of the discretisation. On the one hand one can compute observables for the coarsest discretisation, while on the other hand one can straightforwardly define the continuum limit and return to a description with local degrees of freedom. Indeed, the examples that have been considered so far lead to the conjecture that diffeomorphism symmetry is equivalent to discretisation independence. For quantum mechanical systems (with time discretization), it has been proven in \cite{anharmonic} that diffeomorphism symmetry implies discretization invariance.  Recently, the relationship between diffeomorphism symmetry and discretization independence has been strengthened in the principle of dynamical cylindrical consistency \cite{dittrich-cyl} for time--evolving discrete systems \cite{time-evo}, see also \cite{hoehn}.

This conjecture has been the main motivation of \cite{steinhaus11}: the question has been, whether requiring triangulation independence, i.e. invariance under Pachner moves \cite{pachner}, can be used as a dynamical determination of the path integral measure in (linearized) Regge calculus. One would expect that this requirement would also single out a unique measure. In 3D this lead to a simple measure factor invariant under all Pachner moves, which is consistent with the asymptotics of the Ponzano--Regge (spin foam) model \cite{pr-asymptotics}. The 4D case, that is the expressions for the determinants of the Hessians of the action evaluated on the solution,  turned out to be astonishingly similar to the 3D one, yet these expressions were modified by an overall factor, which appears to be non--factorising with respect to the (sub)simplices of the triangulation and resisted so far a geometric interpretation. More importantly, it has been conjectured in \cite{steinhaus11} to be non--local and, thus, effectively hindering a construction of a  local path integral invariant under a subset of Pachner moves in 4D \cite{steinhaus11}. 

This factor will be the main focus of this paper, in which we will derive a geometric interpretation, namely as a criterion determining whether $d+2$ vertices (in $d$ dimensions) lie on a $(d-1)$--sphere \cite{berger} (see also \cite{dandrea}). This interpretation allows us to prove that this intricate factor is non--factorising, i.e. it cannot be expressed as a simple product of amplitudes associated to the (sub)simplices of the triangulation. (We will refer to this property as `non--local'.) 

This paper is organized as follows: In section \ref{sec:review} we review the setup and main results of \cite{steinhaus11}. Section \ref{sec:interpretation} deals with the derivation of the new interpretation of the non--factorising factor, namely as a criterion determining whether six vertices lie on a 3--sphere, uncovering the factor's non--local nature. To examine the general cases, we have to generalize the study of \cite{steinhaus11} to more general orientations in section \ref{sec:orientation}. We discuss different choices for the measure (see also appendix \ref{app:bary}). The paper is concluded by a discussion of the result in section \ref{sec:discussion}. In appendix \ref{sec:appendix} we discuss some particular cases of the non--local factor.

\section{Linearized Regge calculus and previous results} \label{sec:review}

In the work \cite{steinhaus11} it has been examined whether one can define a triangulation invariant path integral measure for (linearized) length Regge calculus. Let us briefly recap the setup: Consider the Euclidean path integral for the Regge discretization of gravity given on a 3D or a 4D triangulation:
\begin{equation}
\int_{l_e | e \subset \partial M} \prod_e \, d l_e \, \mu(l_e) e^{-S_R[l_e]} \quad ,
\end{equation}
where $l_e | e \subset \partial M$ denotes the fixed edge lengths on the boundary of the triangulated manifold $M$. $S_R[l_e]$ is called the Regge action and is given by the following expression in $d$ dimensions:
\begin{equation} \label{eq:action}
S_R[l_e]:= - \sum_{h \subset \text{bulk}} V_h \, \omega_h^{\text{(bulk)}} 
 - \sum_{h \subset \text{bdry}} V_h \, \omega_h^{\text{(bdry)}} 
 \quad , 
\end{equation}
where $V_h$ is the volume of the $(d-2)$--simplex $h$, also called `hinge', and $\omega^{\text{(bulk)}}_h$ and $\omega^{\text{(bdry)}}_h$ denote the deficit angle or the exterior boundary angle respectively, located at the hinge $h$ in the $d$--dimensional simplex $\sigma^d$. The deficit angles at a hinge are defined as a sum of the dihedral angles of the $d$--simplices sharing the hinge modulo $2 \pi$ and relative orientation of the simplices. The definition for matching orientations of the simplices, also considered in \cite{steinhaus11}, is:
\begin{equation}
\omega^{(\text{bulk})}_h := 2 \pi - \sum_{\sigma^d \supset h} \theta^{(d)}_h \quad ,
\end{equation}
\begin{equation}
\omega^{(\text{bdry})}_h := k \pi - \sum_{\sigma^d \supset h} \theta^{(d)}_h \quad ,
\end{equation}
where $\theta^{(d)}_h$ is the $d$-dimensional dihedral angle at the hinge $h$ in the $d$--simplex $\sigma^d$. $k$ depends on the number of pieces glued together at this boundary. In case only two pieces are put together $k=1$.

This action \eqref{eq:action} has been linearized, i.e. expanded (up to quadratic order) around a flat background solution, denoted by edge lengths $l^{(0)}_e$, that is a solution of the equations of motion $\frac{\partial S_R}{\partial l_e} = 0$ with vanishing deficit angles $\omega^{\text{(bulk)}}_h$ \footnote{In 3D, the solution to the Regge equations of motion (for positive orientation) always implies vanishing deficit angles, however in 4D this is only possible if the boundary data admit a flat solution.}:
\begin{equation}
l_e = l^{(0)}_e + \lambda_e \quad.
\end{equation}
The integration is then performed over the perturbations $\lambda_e$, such that the Hessian matrix of the Regge action, i.e. $\frac{\partial^2 S_R}{\partial l_e \partial l_{e'}}$, becomes the (inverse) `propagator' of the theory. The motivating question of \cite{steinhaus11} has been whether it is possible to define a measure factor $\mu(l^{(0)}_e)$, as a function of the background edge lengths, that allows for a triangulation invariant Regge path integral. To examine triangulation independence, it is enough to only consider local changes of the triangulation, so--called Pachner moves \cite{pachner}: a consecutive application of these transfers a triangulation of a manifold into any other possible triangulation of the same manifold.

In \cite{steinhaus11} exact formulas for the Hessian matrix in 3D and 4D have been derived in great detail from which one concludes a very specific form of measure factors. In the following, we will restrict ourselves to just recalling the main results:

The Hessian matrices one has to compute are of the following form. In 3D we have
\begin{equation}
\frac{\partial^2 S_R}{\partial l_e \partial l_{e'}} = - \frac{\partial \omega_e}{\partial l_{e'}} \quad ,
\end{equation}
where the dihedral angles are associated to the edges. In 4D the situation is more complicated
\begin{equation} \label{eq:4d-second}
\frac{\partial^2 S_R}{\partial l_e \partial l_{e'}} = - \sum_h \frac{\partial A_h}{\partial l_e} \frac{\partial \omega_h}{\partial l_{e'}} - \sum_h \frac{\partial^2 A_h}{\partial l_e \partial l_{e'}} \omega_h - \sum_h \frac{\partial^2 A_h}{\partial l_e \partial l_{e'}} \omega^{\text{(bdry)}}_h \quad ,
\end{equation}
yet only the first term survives, since we are considering a flat background solution $(\omega_h=0)$ and only local changes of the triangulation, which leave the boundary unchanged. Thus for both cases of 3D and 4D Regge calculus, the main task is to compute the first derivatives of the deficit angles.

Although these first derivatives of the deficit angles can be computed explicitly \cite{dittrich07}, it is much more effective to use techniques of \cite{korepanov}, see also \cite{baratin3d}. This utilizes the flat background (for the linearized Regge action) and hence the fact that the  simplicial complex is embeddable into $\mathbb{R}^d$.   One then considers  small deviations in the edge lengths so that the complex remains embeddable,  i.e. the deficit angles $\omega_h$ are unchanged, $\delta \omega_h \equiv 0$.
This requirement automatically translates into a requirement on the variations of the dihedral angles $\delta \theta_h$, which are part of the respective deficit angle. Then one computes the derivatives of the dihedral angles \cite{dittrich07} under the assumption that only two edge lengths are varied at the same time \cite{korepanov}; starting from the simplest case, one derives all other derivatives of the deficit angles by considering the relative change of edge lengths under infinitesimal deviations. Eventually, one finds in 3D: 
\begin{equation} \label{eq:hessian3d}
\frac{\partial^2 S_R}{\partial l_{ij} \partial l_{km}} = - \frac{\partial \omega_{ij}}{\partial l_{km}} = (-1)^{s_i + s_j +s_k +s_m} \frac{l_{ij} l_{km}}{6} \frac{V_{\bar{i}} V_{\bar{j}} V_{\bar{k}} V_{\bar{m}} }{\prod_n V_{\bar{n}}} + \text{bdry terms} \quad ,
\end{equation}
where $l_{ij}$ is the edge lengths between the vertices $i$ and $j$, $\omega_{ij}$ is the deficit angle at the edge $(ij)$ and $V_{\bar{i}}$ is the volume of the tetrahedron obtained by removing the vertex $i$ \footnote{This assignment is unique in subsets of the triangulation, which are subject to a Pachner move. These subsets consist of $d+2$ vertices in $d$ dimensions; by removing one vertex, $d+1$ vertices remain, which span a $d$--simplex.}. The signs $s_i$ depend on the orientation and the considered Pachner move. We provide a different derivation of them in section \ref{sec:interpretation}. In this work we will neglect the boundary terms, since they are not relevant for the main argument of this paper, yet they are essential to show that the classical Regge action is invariant under Pachner moves \cite{steinhaus11}.

Indeed, the idea to construct a triangulation invariant measure factor, relies crucially on the invariance of the classical Regge action under Pachner moves. In 3D this is the case for all Pachner moves, such that we were able to derive an invariant measure factor, which is factorising, and hence local, with a straightforward geometrical interpretation:
\begin{equation}
\mu(\{l\}) = \frac{\prod_e \frac{1}{\sqrt{12 \pi}} \, l_e}{\prod_\tau \sqrt{ V_\tau}} \quad.
\end{equation}
To each edge $e$ of the triangulation one associates the edge length $l_e$ and a numerical factor of $(12 \pi)^{-\frac{1}{2}}$, to each tetrahedron $\tau$ one associates the inverse (square root) of its volume $V_\tau$. This measure factor is consistent, even up to the numerical factor\footnote{The association of the factor $(\sqrt{12 \pi})^{-1}$ is not unambiguous. It can either be assigned to edges or tetrahedra.}, with the asymptotic expansion of the $\text{SU}(2)$ $6j$ symbol \cite{pr-asymptotics}, the amplitude associated to a tetrahedron in the Ponzano Regge model, which is triangulation invariant (and topological) as well.

In 4D Regge calculus the situation is more complicated, as one might expect: First the Regge action itself is in general not invariant under the 3--3 Pachner move \cite{steinhaus11}. It is a peculiar move, since both possible configurations only differ by the triangle shared by all three 4--simplices of this configuration, no dynamical edge is involved. Hence the configurations are solely determined by the boundary data, which might be chosen such that the deficit angle on the bulk triangle does not vanish, i.e. the configuration is not flat. However as soon as this is the case, the Regge action is not invariant under this Pachner move any more. Thus the Regge action is not a suitable starting point to define an invariant measure under all 4D Pachner moves.

Nevertheless the 4D Regge action is invariant under the 5--1 and 4--2 Pachner moves (and their inverses), so it has been examined whether one can define a measure factor that is at least invariant under these two local changes of the triangulation. Surprisingly, the (considered part of the) Hessian matrix is very similar to the 3D case:
\begin{equation} \label{eq:hessian4d}
\frac{\partial^2 S_R}{\partial l_{op} \partial l_{mn}} = - \sum_{k \neq o,p} \frac{\partial A_{opk}}{\partial l_{op}} \frac{\partial \omega_{opk}}{\partial l_{mn}} = D_{op} \, (-1)^{s_o + s_p + s_m + s_n} \frac{l_{op} l_{mn}}{96} \frac{V_{\bar{o}} V_{\bar{p}} V_{\bar{m}} V_{\bar{n}}}{\prod_{l} V_{\bar{l}}} + \text{bdry terms} \quad ,
\end{equation}
where the factor $D_{op}$ is the only difference to the 3D case, besides the fact that the $V_{\bar{i}}$ now denote volumes of 4--simplices. Note that for the 5--1 move in 4D (and similarly the 4--1 move in 3D), the Hessian matrix possesses four null eigenvectors, which correspond to a vertex translation invariance of the subdividing vertex. The divergent part of the integral is then identified as an integral over a 4D volume and is gauge fixed to 1. We also provide a brief derivation of \eqref{eq:hessian4d} in section \ref{sec:orientation} for more general orientations.

The emphasis of this paper lies on this factor $D_{op}$, which is given by the following expression:
\begin{equation} \label{eq:factor-D}
D_{op} := \sum_{k\neq o,p} (-1)^{s_k} \left(l_{ok}^2 + l_{pk}^2 - l_{op}^2 \right) V_{\bar{k}} \quad .
\end{equation}
Since the Hessian matrix is symmetric, one concludes that $D_{op}$ actually does not depend on the vertices $o$ and $p$, such that it turns into an overall factor of the Hessian matrix. Thus we will only refer to it as $D$ in the following.

Ignoring $D$ for the time being, one can construct an `almost' triangulation invariant measure, very similar to the 3D case:
\begin{equation} \label{eq:nonlocal}
\mu(\{l\}) = \frac{\prod_e \frac{1}{\sqrt{192 \pi}} \, l_e}{\prod_\Delta \sqrt{ V_\Delta}}\quad ,
\end{equation}
where $V_\Delta$ now denotes the volume of the 4--simplex $\Delta$. The numerical factors are again assigned to the edges\footnote{As in 3D, the assignment of numerical factors is not unambiguous.}.

Despite the concise expression \eqref{eq:factor-D} of $D$, it both impedes the construction of a triangulation invariant measure and resists a nice geometric interpretation, since it is not obvious, whether it can be written as a product of amplitudes associated to (sub)simplices. Hence it has been conjectured in \cite{steinhaus11} that $D$ is non--local and cannot be written in a factorising way. The purpose of this paper is to provide a  geometric interpretation for the factor $D$, namely it is a criterion that determines whether the 6 vertices, that make up the simplicial complex (in 4D) to which the Pachner move is applied, lie on a 3--sphere. We will use this to prove that the factor $D$ generically cannot be accommodated by a local measure factor, in particular not by simple product (or quotient) of volumes of (sub)simplices.

\section{A geometric interpretation for $D$} \label{sec:interpretation}

In this section we will derive a geometric interpretation for the factor $D$ mentioned above. Actually, the definition \eqref{eq:factor-D} of $D$ is valid in any dimension $d \geq 3$, such that we will derive its geometric interpretation for arbitrary dimensions\footnote{Note that it is not clear whether $D$ will also arise in higher dimensions in the framework of linearized Regge calculus. We can only confirm this for $d=4$ \cite{steinhaus11}.}. 

In the following we will discuss a simplicial complex in $d$ dimensions, to which a Pachner move will be applied. Consider $d+2$ vertices embedded in $\mathbb{R}^d$, such that they form non--degenerate $d$--simplices. The geometry can be completely characterized by the set of the edge lengths $\{l_{ij}\}$, describing the Euclidean distances between the vertices. Given a set of vertices and the edge lengths between them, one can define the associated Cayley--Menger matrix $C$ \cite{berger,sorkin,korepanov}. In the case we consider here, this is a $(d+3)\times(d+3)$--dim. matrix given by:
\begin{equation} \label{eq:cayley-menger}
C := \left(
\begin{matrix}
0 & 1 & 1 & 1 & \cdots & 1 \\
1 & 0 & l_{01}^2 & l_{02}^2 & \cdots & l_{0(d+1)}^2 \\
1 & l_{01}^2 & 0 & l_{12}^2 & \cdots & l_{1(d+1)}^2 \\
\vdots & \vdots & \vdots & \vdots & \ddots & \vdots \\
1 & l_{0(d+1)}^2 & l_{1(d+1)}^2 & l_{2(d+1)}^2 & \cdots & 0
\end{matrix}
\right) \quad .
\end{equation}
In general, the determinant of the Cayley--Menger matrix, $\det C$, associated to a $d$--simplex is proportional to the square of its $d$--volume. However, in the example at hand, the $d+2$ vertices are embedded in $\mathbb{R}^d$, such that they form a degenerate $(d+1)$--simplex, hence $\det C=0$. Since we have required that the $d$--simplices are non--degenerate, $C$ has exactly one null eigenvector, corresponding to changes of the edge lengths, such that the $d+2$ vertices remain (embeddable) in $\mathbb{R}^d$. To describe this null eigenvector, let us introduce some notation.

By $C^i_j$ we denote the submatrix of $C$ obtained by deleting its $i$'th column and its $j$'th row with $i,j \, \in \{0,1,2,\dots,d+2\}$. The determinant of the submatrix, i.e. the $(i,j)$th minor of $C$, is denoted by $\left| C^i_j \right| := \det C^i_j$. To simplify notation we simply call the diagonal minors $\left|C_i\right|$. In fact, since $\det C = 0$, all off-diagonal minors can be expressed in terms of the diagonal ones \cite{kokkendorff}:
\begin{equation}
\left|C^i_j\right| = \sqrt{\left|C_i\right|}\sqrt{\left|C_j\right|} \quad .
\end{equation}
Before we construct the null eigenvector of $C$, it is instructive to examine the minors $C_0$ and $C_i$ for $i >0$ in more detail and discuss their geometric interpretation.

\begin{figure}[h!]
\begin{center}
\includegraphics[scale=0.8]{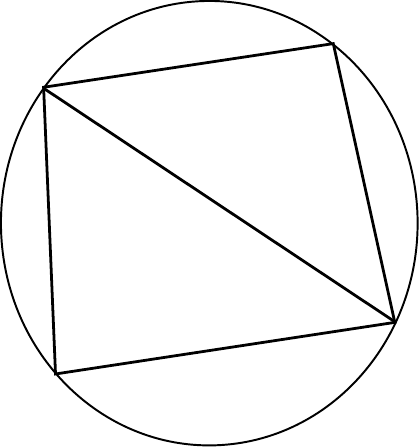}
\caption{\label{fig:2d}
The situation in 2D: Four vertices on a 1--sphere.}
\end{center}
\end{figure}

The matrix $C_0$, obtained by deleting the `0'th column and row of $C$ is particularly important in this paper. It is given by:
\begin{equation} \label{eq:D-Cayley}
C_0 = \left(
\begin{matrix}
0 & (l_{01})^2 & \dots & (l_{0(d+1)})^2 \\
(l_{01})^2 & 0 & \dots & (l_{1(d+1)})^2 \\
\vdots & \vdots & \dots & \vdots \\
(l_{0(d+1)})^2 & (l_{1(d+1)})^2 & \dots & 0
\end{matrix}
\right)  \quad .
\end{equation}
It has been shown in \cite{berger} (see also \cite{dandrea}) that $|C_0|$ has a very specific geometric meaning. In case $|C_0|=0$ the $d+2$ vertices lie on a $(d-1)$--dimensional sphere, see fig. \ref{fig:2d} for an example in $d=2$. In a sense, this is a non--local statement, since it can only be deduced if the positions of all $d+2$ vertices are known; it cannot be inferred from just $d+1$ vertices. From the construction of the null eigenvector, we will show that $D \sim \sqrt{|C_0|}$. Hence we argue that $D$ is non--local in section \ref{sec:non-local}.

The second interesting minor we want to discuss is $|C_i|$ for $i \in \{1,2,\dots,d+1\}$. If one takes a look at the definition of $C$ in \eqref{eq:cayley-menger} again, then one realizes that by removing the $i$'th row and column, one removes all edge lengths with the index $i-1$. The remaining matrix is again a Cayley-Menger matrix, yet for a $d$--simplex. Hence 
\begin{equation}
|C_i| = (-1)^{d+1} 2^d (d!)^2 \left(V_{\overline{i-1}}\right)^2 \quad .
\end{equation}

The single null eigenvector of $C$ is given by\footnote{The null eigenvector can be deduced from the expansion of $\det C$ with respect to different rows.} 
\begin{equation} \label{eq:null-ev}
\vec{v}_c := \left((-1)^s\sqrt{|C_0|},(-1)^{s_0} \sqrt{|C_1|},\ldots,(-1)^{s_{d+1}} \sqrt{|C_{d+2}|}\right) = \left( (-1)^s\sqrt{|C_0|},(-1)^{s_0} \, 2^{d/2} d! \, V_{\bar{0}},\ldots,(-1)^{s_{d+1}} \, 2^{d/2} d! \, V_{\overline{d+1}} \right) \quad .
\end{equation}
The signs $s$ and $s_i$ are determined only up to an overall sign ambiguity. We will show below that we can choose $s_i$ to be the same signs that appear also in \eqref{eq:hessian4d} and \eqref{eq:factor-D}. The action of the matrix $C$ on $\vec{v}_c$ gives the following relations:
\begin{align} \label{eq:null-ev-eq}
\sum_{i=0}^{d+1} (-1)^{s_i} V_{\bar{i}} & = 0 \quad , \\
\forall_{j\geq 0} \quad \quad (-1)^s\sqrt{|C_0|} + \sum_{i = 0}^{d+1} (-1)^{s_i} \, 2^{d/2} d! \, l_{ij}^2V_{\bar{i}} & =0 \quad . \label{eq:null-ev-eq2}
\end{align} 
The first condition fixes the relative signs $s_i$ and is straightforward to interpret if one recalls the change of the triangulation under the Pachner move. Take a $1-(d+1)$ move for example, where the new vertex $0$ is added inside to the $d+1$ vertices forming the $d$--simplex. The relation shows that the volume before and after the Pachner move is the same:
\begin{equation}
V_{\bar{0}} = \sum_{i=1}^{d+1} V_{\bar{i}} \quad,
\end{equation}
which fixes the signs $s_0=1$ and $s_i=0 \, \forall \, i \geq 1$, consistent with the results of \cite{steinhaus11}. Similar relations also hold for the other Pachner moves. For different (relative) orientations (see section \ref{sec:orientation}), one or more signs get flipped such that such a clear separation between volumes `before' and `after' the Pachner move is not possible any more. In fact, since we mainly discuss the 5--1 move in this paper, we fix $s_0=1$, i.e. the orientation of the coarse simplex, because it is determined by the boundary data and is thus not affected by moving the vertex $0$. This singles out the coarse simplex as a particular reference frame with respect to which the relative orientation of the other simplices is defined.

At this point we can also determine the sign $s$ using equation \eqref{eq:null-ev-eq2} for $j=0$
\begin{equation}
 (-1)^s\sqrt{|C_0|} = - 2^{d/2} d! \, \sum_{i = 1}^{d+1} (-1)^{s_i} l_{i0}^2V_{\bar{i}} =- 2^{d/2} d! \, \sum_{i = 1}^{d+1} l_{i0}^2V_{\bar{i}}<0  \quad .
\end{equation}
Thus we conclude $s=1$.

The relations \eqref{eq:null-ev-eq} can be used to derive a relation between the non--local factor $D$ and the criterion $|C_0|$ determining whether $(d+2)$ vertices lie on a $(d-1)$--sphere. Recall that for arbitrary $i \neq j$, $D$ was defined as:
\begin{equation}
 D_{ij}=\sum_{k\neq i,j} (-1)^{s_k} (l_{ik}^2 + l_{jk}^2 - l_{ij}^2) V_{\bar{k}} \quad .
\end{equation}
Let us expand $D_{ij}$ as follows:
\begin{equation} \label{eq:main}
 D_{ij}=\underbrace{\sum_{k} (-1)^{s_k} (l_{ik}^2 + l_{jk}^2) V_{\bar{k}}}_{\overset{\eqref{eq:null-ev-eq}}{=} -2 (-1)^s  2^{-d/2} (d!)^{-1} \sqrt{|C_0|}} \underbrace{- \left( (-1)^{s_j} l_{ij} V_{\bar{j}} + (-1)^{s_i} l_{ij} V_{\bar{i}} \right) - \sum_{k \neq i,j} (-1)^{s_k} l_{ij}^2 V_{\bar{k}}}_{=-l_{ij}^2 \sum_k (-1)^{s_k} V_{\bar{k}} \overset{\eqref{eq:null-ev-eq}}{=} 0} = - 2^{1-d/2} (d!)^{-1} (-1)^s\sqrt{|C_0|} \quad .
\end{equation}
This proves the relation between the non--local measure $D$ and the criterion $|C_0|$.

This automatically gives a new interpretation to the non--factorising factor $D \sim \sqrt{|C_0|}$ appearing in the 4D Pachner moves. If it vanishes\footnote{The result is that the quadratic part of the action vanishes and the integral diverges.} all six vertices of the 4--simplices involved in the Pachner move lie on a 3--sphere. To determine whether this is the case or not, one has to know the positions of all six vertices with respect to each other, it cannot be inferred from a subset. Thus it is already implied that the factor $D$ has to be non--local, since its geometric meaning can only be deduced if the relative positions of all six vertices are known. We will use this fact in section \ref{sec:non-local} to show that $D$ does not factorise.


This geometric interpretation is even more pronounced if we express the factor $D$ in affine coordinates. To this end we specialize to the $(d+1)-1$ move, in which we integrate out $d$ edge lengths, that start from a subdividing vertex $0$, which lies inside the final simplex.

An efficient way to describe the coordinate of the subdividing vertex 0 with respect to the final simplex $\bar{0}$ is by using affine coordinates. The idea is to write the position vector $\vec{x}_0$ of the new vertex as a weighted sum of the position vectors $\vec{x}_i, i \neq 0$, with weights $\alpha_i$. The condition $\sum_{i\not=0} \alpha_i=1$ ensures that this prescription is well--defined. If one additionally requires that $\alpha_i \geq 0, \; \forall i \neq 0$, then the new vertex is inside the final simplex. As soon as one of the $\alpha_i$ is negative, the vertex 0 is located outside.

Hence, the position vector $\vec{x}_0$ is given by
\begin{equation}
 \vec{x}_0=\sum_{i\not=0} \alpha_i \vec{x}_i \quad ,
\end{equation}
thus
\begin{equation}
 \vec{x}_k-\vec{x}_0= \left(\sum_{i \neq 0} \alpha_i\right) \vec{x}_k - \sum_{i \neq 0} \alpha_i \vec{x}_i = \sum_{i \neq 0} \alpha_i (\vec{x}_k - \vec{x}_i) \quad .
\end{equation}
The (square of the) new edge lengths is given by $l_{0k}^2=(\vec{x}_k-\vec{x}_0)^2$:
\begin{align}
 l_{0k}^2 & =\sum_{i\not=0,j\not=0} \alpha_i\alpha_j(\vec{x}_k-\vec{x}_i) \cdot(\vec{x}_k-\vec{x}_j)= \frac{1}{2} \sum_{i \neq 0,j\neq 0} \alpha_i \alpha_j \left( (\vec{x}_k-\vec{x}_i)^2 + (\vec{x}_k - \vec{x}_j)^2 - \left( (\vec{x}_k - \vec{x}_i) - (\vec{x}_k- \vec{x}_j) \right)^2 \right) \nonumber \\
  & = \frac{1}{2}\sum_{i\not=0,j\not=0} \alpha_i\alpha_j \left(l_{ik}^2+l_{jk}^2-l_{ij}^2 \right)= \sum_{i \neq 0} \alpha_i l_{ik}^2 \underbrace{\left( \sum_{j \neq 0} \alpha_j \right)}_{=1}  -\underbrace{\sum_{0<i<j} \alpha_i\alpha_jl_{ij}^2}_{=:b^2} =\sum_{i \neq 0} \alpha_i l_{ik}^2 - b^2 \quad .
\end{align}
Recalling the determinant expression of $D^2$, i.e. \eqref{eq:main} and \eqref{eq:D-Cayley}, $l_{0k}^2$ are the entries of the first row and column of the matrix $C_0$. Without changing the determinant we subtract $\alpha_i$ times the $(i+1)$'th row from the first one for all $i\neq 0$ and obtain:
\begin{equation}
 D^2= \left( \frac{1}{48} \right)^2 \det\left(
 \begin{array}{ccccc}
 -b^2 & -b^2 & -b^2 & \cdots & -b^2 \\
  \sum_{i\not=0} \alpha_il_{i1}^2-b^2 & 0 & l_{12}^2 &\cdots & l_{15}^2\\
  \sum_{i\not=0} \alpha_il_{i2}^2-b^2 & l_{12}^2 & 0 & \cdots & l_{25}^2\\
  \vdots & \vdots & \vdots &\ddots &\vdots\\
  \sum_{i\not=0} \alpha_il_{i5}^2-b^2 & l_{15}^2 & l_{25}^2 &\cdots &0
 \end{array}\right) \quad ,
\end{equation}
where it is straightforward to derive that $\sum_{j\not=0}\alpha_j \left(\sum_{i\not=0} \alpha_il_{ij}^2-b^2 \right)=b^2$.
Again, we repeat the procedure for the columns by subtracting $\alpha_i$ times the $(i+1)$'th column from the first one for all $i \neq 0$. Then the result can be written as
\begin{equation} \label{eq:D-affine}
 D^2= \left( \frac{1}{48} \right)^2 \det\left(
 \begin{array}{ccccc}
 0& -b^2 & -b^2 & \cdots & -b^2 \\
  -b^2 & 0 & l_{12}^2 &\cdots & l_{15}^2\\
  -b^2 & l_{12}^2 & 0 & \cdots & l_{25}^2\\
  \vdots & \vdots & \vdots &\ddots &\vdots\\
  -b^2 & l_{15}^2 & l_{25}^2 &\cdots &0
 \end{array}\right)= 4 b^4 \, V_{\bar{0}}^2 = 4 \left(\sum_{0<i<j}\alpha_i\alpha_j l_{ij}^2\right)^2 V_{\bar{0}}^2 \quad .
\end{equation}
Indeed, \eqref{eq:D-affine} is a remarkable identity for $D$ (in the 5--1 move\footnote{ As discussed in section \ref{sec:orientation}, it is also possible to describe $D$ in a similar way for the 4--2 move. However, there one has to choose one of the two 4--simplices (in the 2 simplex configuration) as a reference frame to describe the fixed position of the additional vertex. Clearly, this is not unambiguous.}): The non--locality of $D$ here is encoded in the choice of reference frame, namely the final simplex. In particular from the perspective of the five simplex configuration, this is not obvious. Moreover, the dependence on the position of the new vertex only enters into the factor $b^2$ through the weights $\alpha_i$. 

Let us describe the geometric meaning of $b^2 = \sum_{0<i<j}\alpha_i\alpha_j l_{ij}^2 $. Let $\vec{x}$ denote the position vector of the circumcenter of the $4$--simplex $\bar{0}$, i.e. the circumcenter of the 3--sphere circumscribing the 4--simplex. Then the distance between this point and the vertex 0 is given by:
\begin{equation}
(\vec{x}_0-\vec{x})^2=\sum_{i\not=0, j\not=0} \alpha_i\alpha_j (\vec{x}_i-\vec{x})(\vec{x}_j-\vec{x})=
\frac{1}{2}\sum_{i\not=0, j\not=0} \alpha_i\alpha_j \left((\vec{x}_i-\vec{x})^2+(\vec{x}_i-\vec{x})^2-(\vec{x}_i-\vec{x}_j)^2\right) \quad ,
\end{equation}
yet, by definition of $\vec{x}$, $(\vec{x}_i-\vec{x})^2=r^2 \; \forall i \neq 0$. Thus
\begin{equation}
 (\vec{x}_0-\vec{x})^2=r^2-\sum_{0<i<j} \alpha_i\alpha_jl_{ij}^2
\end{equation}
and finally 
\begin{equation} \label{eq:b-radius}
b^2=r^2-(\vec{x}_0-\vec{x})^2 \quad .
\end{equation}
From this fact and \eqref{eq:D-affine} we can simply deduce several properties of $D$: First of all, since $V_{\bar{0}}\not=0$ (by assumption), we conclude that $D=0$ exactly if $b^2 = 0$. From identity \eqref{eq:b-radius} it is clear that this happens only if the vertex 0 lies on the circumscribing sphere of the 4--simplex $\bar{0}$, namely $(\vec{x} - \vec{x}_0)^2 = r^2$. On the other hand, we can directly consider the definition of $b^2$:
\begin{equation} \label{eq:def-b}
b^2 = \sum_{0<i<j} \alpha_i \alpha_j l_{ij}^2 \quad .
\end{equation}
It is straightforward to recognize that if we restrict the vertex 0 to stay inside the 4--simplex, i.e. $\alpha_i \geq 0\; \forall i$, then $b^2=0$ is only possible if all $\alpha_i$ except one vanish.

\begin{figure}[h!]
\begin{center}
\includegraphics[scale=0.8]{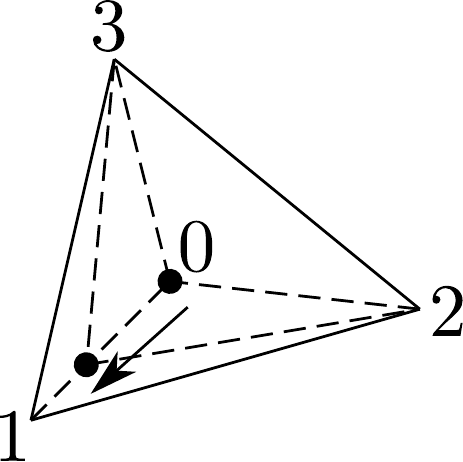}
\caption{ \label{fig:degenerate} 
The degenerate limit in $2$D, in which the inner vertex $0$ is moved on top of one of the vertices, here $1$, of the original triangle. Clearly, once the limit is reached the areas of the triangles $(012)$ and $(013)$ vanish.}
\end{center}
\end{figure}

This is exactly the case when the vertex $0$ is moved on top one of the vertices of the final (coarser) simplex. See fig. \ref{fig:degenerate} for the 2D case.   In 4D four of the initial five simplices become degenerate and from the limits 
\begin{equation}
\lim_{(0)\rightarrow (1)} l_{01} = 0 \quad , \lim_{(0) \rightarrow (1)} l_{0k} = l_{1k}, \; \forall k \neq 0,1 \quad ,
\end{equation}
\begin{equation}
\lim_{(0) \rightarrow (1)} V_{\bar{1}} = V_{\bar{0}} \quad , \lim_{(0) \rightarrow (1)} V_{\bar{k}} = 0, \; \forall k \neq 0,1 \quad ,
\end{equation}
for the volumes and the length variables it is clear that $\lim_{(0) \rightarrow (1)} D = 0$.   This limit, in which one vertex is moved on top of another plays a crucial role in the relation between diffeomorphism symmetry and triangulation independence \cite{anharmonic}: in the discrete diffeomorphism symmetry is realized as an invariance with respect to moving vertices. Indeed finding a triangulation invariant measure for the $5-1$ move would also imply invariance of the path integral under changing the position of the subdividing vertex $0$.  The classical action (or rather Hamilton--Jacobi functional, i.e. the action evaluated on the solution) already has this symmetry.

An extreme case is given by moving vertices to on top of other vertices, which effectively coarse grains the triangulation. One would  expect a singular behaviour in this case, as  the lengths variables one integrates over become redundant (we are in the linearized theory, and moving the vertex actually affects the background variables). 

Apart from these degenerate cases it can happen that $D$ vanishes, if we move the subdividing vertex outside the coarser simplex. In this case some  some $\alpha_i < 0$.  Indeed, the conditions $b^2 = 0$ and $\sum_i \alpha_i = 1$ fix three of the five $\alpha_i$, parametrising a 3--sphere, thus explaining the previous geometric interpretation.
Yet these cases involve a change of orientation, which will be reflected in the definition of the deficit angle and hence the action. One might be concerned that the formulas derived in \cite{steinhaus11} are no longer valid, but we will show in section \ref{sec:orientation} that the same arguments work also in this case.

\subsection{Non--locality of the measure} \label{sec:non-local}

The geometric interpretation of the factor $D$ reveals its non--local structure, that it does not factorize over \mbox{(sub)simplices} of the triangulation.  
Assuming that it would factorize, i.e. $D$ is of the form
\begin{equation}\label{w1}
D= \prod_\sigma A_\sigma(l)\quad ,
\end{equation}
where the product is over the 4--simplices  and all the other  lower dimensional simplices in the complex under consideration.  Note that factors in (\ref{w1}) are allowed to be constant.  A factor $A_\sigma(l)$ depends only on the length variables of the edges contained in this simplex (which completely specify the geometry of this simplex). 
The zeros of $D$ would be given by the union of the zeros of all its subfactors. However each subfactor can only depend  on the lengths of the edges connecting these five vertices of the simplicial complex in question (which has six vertices), whereas the vanishing of $D$ is equivalent to a condition involving all six vertices.

Assume that $D$, and hence at least one of the factors in (\ref{w1}) is vanishing.  Thus the six vertices of the simplicial complex in question lie on one 3--sphere. Choose one of the vanishing factors, say $A_{\sigma'}$. As $\sigma'$ does not include all vertices, we can change the position of one of the vertices not in $\sigma'$ so that the six vertices do not lie on a 3--sphere any more. In this case $D$ still vanishes due to the factorizing nature assumed in (\ref{w1}), which contradicts that $D$ is only vanishing if the six vertices are on the 3--sphere. Thus $D$ cannot be of the factorizing form (\ref{w1}).

\section{Orientation} \label{sec:orientation}

In this section we will discuss the changes of orientation that occur once the additional vertex is moved outside the coarser $4$--simplex. As we will explain below, this will only result in the change of certain signs; the entire derivation performed in \cite{steinhaus11} works analogously.

Before we move the vertex outside, let us first consider how the null eigenvector $\vec{v}_c$ of $C$ (see eq. \eqref{eq:null-ev}) is affected, when we move the vertex toward the boundary of the coarser simplex along a straight line. As we will see, one or more entries of $\vec{v}_c$ will vanish once the boundary is reached; which ones and how many depends on the dimension of the subsimplex the internal vertex is placed upon. The vanishing entries will change sign once the boundary is crossed (through this subsimplex), which corresponds to a change of orientation of the $d$--simplices, which share the before mentioned subsimplex.

\begin{figure}[h!]
\begin{center}
\includegraphics[scale=0.75]{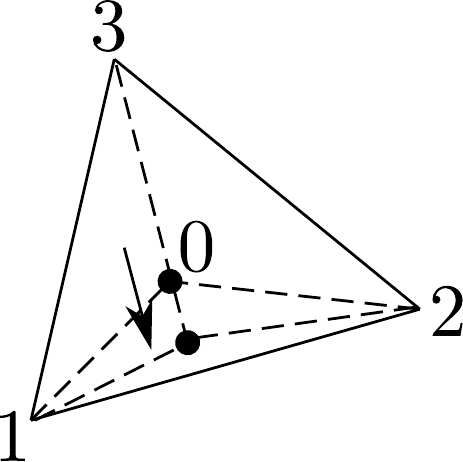}
\caption{\label{fig:2d-sign}
Approaching the boundary of the triangle $(123)$ with vertex $0$ via the edge $(12)$. If the vertex $0$ reaches the edge, the area of the triangle $(012)$, i.e. $V_{\bar{3}}$, vanishes.}
\end{center}
\end{figure}

To illustrate this point, let us revisit the simple 2D example. Consider again the triangle $(123)$ spanned by three vertices, which is subdivided in a 1--3 move by the vertex $0$ placed in its center. If one intends to move the vertex $0$ outside the triangle $(123)$, one has two options: either one crosses through an edge or a vertex, i.e. a 1--dim. or a 0--dim. subsimplex. If one moves $0$ closer to the edge $(12)$ as illustrated in fig. \ref{fig:2d-sign}, the volume of the triangle $(012)$, $V_{\bar{3}}\rightarrow 0$. Once the vertex is moved across the boundary, the component $\sim V_{\bar{3}}$ in $\vec{v}_c$ changes its sign, here $s_3$.

The other situation, when the vertex $0$ is moved on top of one of the other vertices has been already discussed in section \ref{sec:interpretation} and illustrated in figure \ref{fig:degenerate}: The vertex $0$ is moved towards the vertex $1$, such that, once they are on top of each other, the areas of the triangles $(012)$ and $(013)$ vanish, i.e. $V_{\bar{2}} = V_{\bar{3}} =0$. Additionally, since all four vertices lie on the same 1--sphere, $C_0=0$. Hence, three signs in $\vec{v}_c$ change, namely $s$, $s_2$ and $s_3$.

\begin{figure}[h!]
\begin{center}
\includegraphics[scale=0.65]{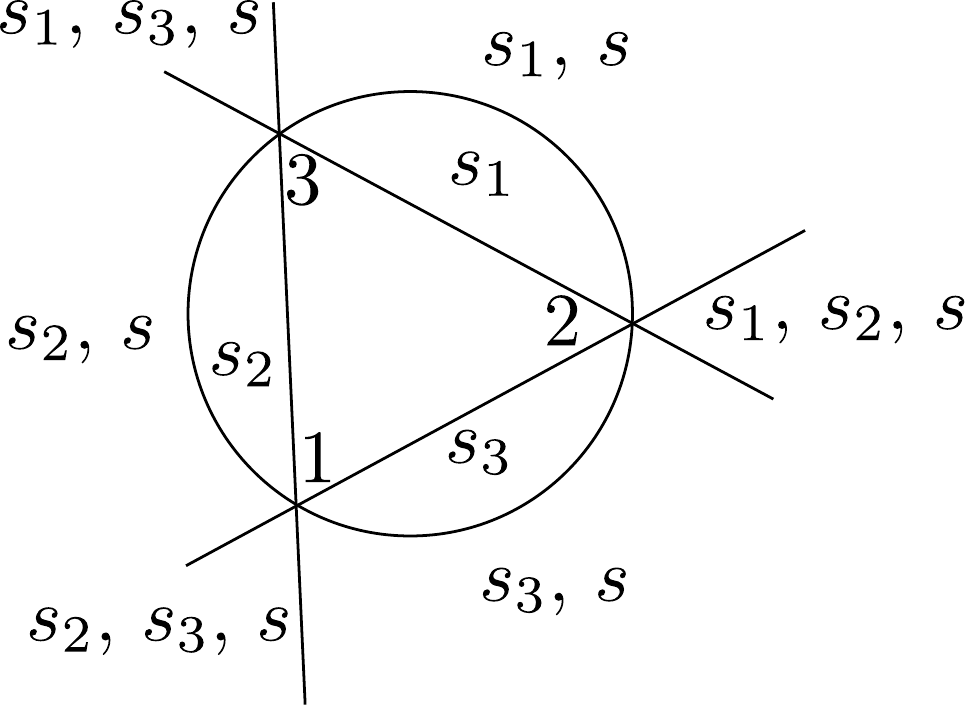}
\caption{\label{fig:signs2D}
The triangle $(123)$ and the 1--dim. hypersurfaces determining the change of sign. A region with $s_i$ inside denotes that the sign $s_i$ is changed, if the vertex is moved $0$ is moved from inside the triangle into this labelled region.}
\end{center}
\end{figure}

It is straightforward to generalize these ideas to arbitrary dimensions $d$: The relative signs $s$, $s_i$ depend on the position of the subdividing vertex $0$ with respect to $(d-1)$--dim. hypersurfaces: In the case of the sign $s$, it is the $(d-1)$--sphere circumscribing the initial $d$--simplex. For the sign $s_i$, it is the hypersurface orthogonal to the normal vector associated to one $(d-1)$--simplex formed without the vertex $i$. We illustrate this in 2D in fig. \ref{fig:signs2D}.

To be more concrete, let us examine all possible situations in 4D: Consider the initial 4--simplex $(12345)$ formed by five vertices. The vertex $0$ can `leave' the initial 4--simplex either through a tetrahedron, a triangle, an edge or a vertex. We have summarized the respective changes of signs in the following table:
\begin{center}
\begin{tabular}{|l|l|l|}
\hline
 Approached / crossed subsimplex: & Vanishing entries of $\vec{v}_c$ & Changes of signs\\
 \hline
 Tetrahedron $(1234)$ & $V_{\bar{5}}=0$ & $s_5$\\
 Triangle $(123)$ & $V_{\bar{4}}=V_{\bar{5}}=0$ & $s_4$, $s_5$\\
 Edge $(12)$ & $V_{\bar{3}}=V_{\bar{4}}=V_{\bar{5}}=0$ & $s_3$, $s_4$, $s_5$\\
 Vertex $1$ & $C_0=V_{\bar{2}}=\cdots=V_{\bar{5}}=0$ & $s$ and  $s_2,\cdots s_5$\\
 \hline
\end{tabular}
\end{center}

\begin{figure}[h!]
\begin{center}
\includegraphics[scale=0.65]{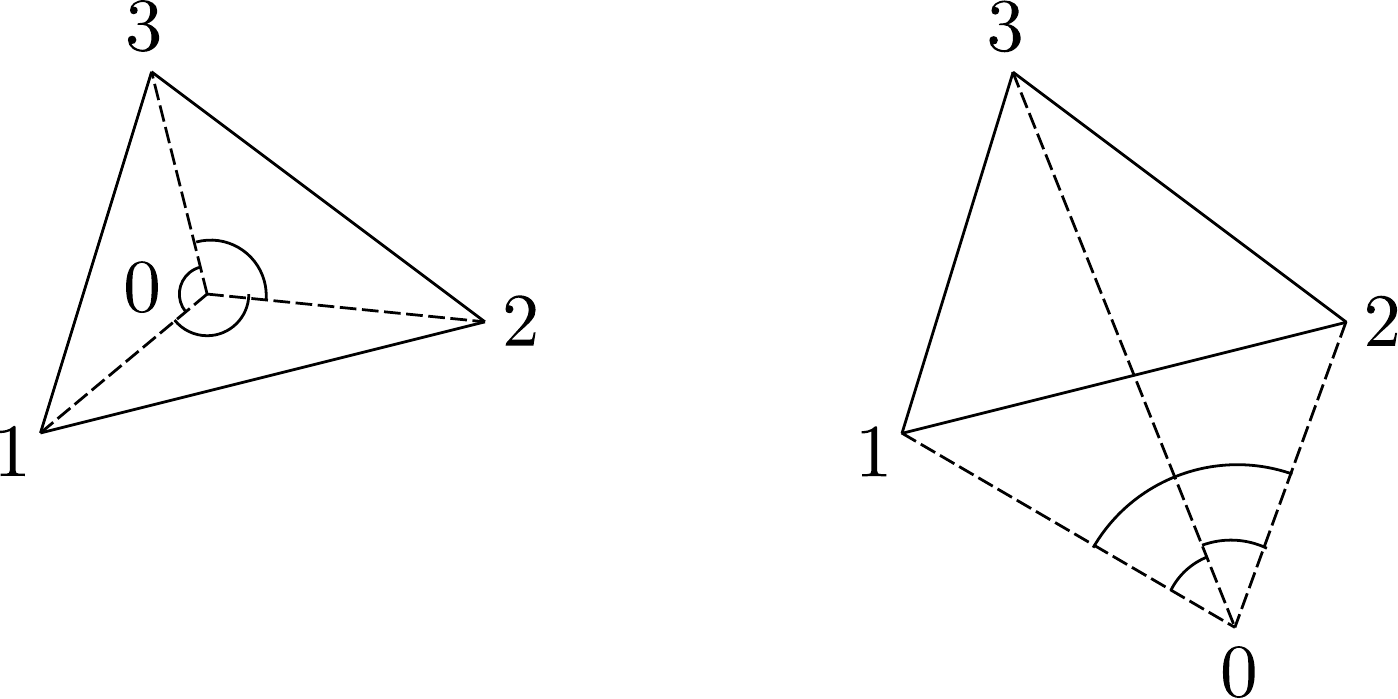}
\caption{\label{fig:orientation} We consider here the 1--3 Pachner, in which the triangle $(123)$, formed by the vertices 1, 2 and 3, is subdivided into three triangles $(012)$, $(013)$ and $(023)$ by adding a new vertex $0$ and connecting it to all old vertices. On the left the `usual' configuration after the 1--3 Pachner move with the vertex $0$ inside the triangle is depicted, on the right the vertex 0 is outside the triangle, which results here in the opposite orientation of the triangle $(012)$.}
\end{center}
\end{figure}

Let us now focus on the change of deficit angles in the action that occurs when we cross the boundary of the initial simplex.
To explain our point and provide an intuitive example, let us consider the two different orientations drawn in fig. \ref{fig:orientation} for the 1--3 Pachner move in $2$D. On the left, we show the well--known configuration, in which the original triangle gets subdivided by an additional vertex $0$ into the three triangles $(012)$, $(013)$ and $(023)$. There are three triangles meeting at this new vertex and in order to form a flat triangulation, the dihedral angles located at vertex $0$, called $\theta_0^{(0ij)}$ for $i,j \in \{1,2,3\}$ with $i\neq j$, have to sum up to $2\pi$. On the right, the same new triangles are created by adding the vertex $0$, but this new vertex is now located outside the triangle, which corresponds to a change of orientation. Let us discuss this further.

To be more precise, moving the vertex $0$ outside the triangle $(123)$ through the edge $(12)$ changes the relative orientation of the triangle $(012)$ with respect to the triangles $(013)$ and $(023)$. This results into one sign swap in the definition of the deficit angle at the vertex $0$ (modulo $2\pi$):
\begin{equation} \label{eq:deficit-or}
\omega_0 = \theta_0^{(013)} + \theta_0^{(023)} - \theta_0^{(012)} = \sum_t \epsilon_t \theta_0^t \quad .
\end{equation}
The last part of equation \eqref{eq:deficit-or} denotes the formal sum over all triangles $t$ meeting at the vertex $0$, where each triangle now carries a colouring $\epsilon_t \in \{\pm 1 \}$ denoting its relative orientation, see also \cite{baratin3d}. This simplicial complex is embeddable into $\mathbb{R}^2$, if the deficit angle $\omega_0$ at vertex $0$ vanishes modulo $2 \pi$\footnote{Due to the peculiar relative orientation of simplices one can argue \cite{antispacetime} that the curvature does not vanish at this vertex.}. This condition can be automatically translated into the fact that the dihedral angles (at vertex $0$) in the triangles $(013)$ and $(023)$ have to sum up to the one in the triangle $(012)$, which can be nicely seen in fig. \ref{fig:orientation}.
 
Let us return to the 4D case. As the argument above shows, the change of relative signs in the definition of the deficit angle coincides with the change of relative signs of the 4--simplices, i.e. the signs $s_k$. From that we deduce
\begin{equation}
 \epsilon_{\bar{k}} = (-1)^{s_k} \quad .
\end{equation}
Finally the general definition of the deficit angle (modulo $2\pi$) located at the triangle $(ijk)$ is:
\begin{equation}
 \omega_{ijk}=\sum_l (-1)^{s_l} \theta^{\bar{l}}_{ijk} \text{ mod } 2\pi \quad  .
\end{equation}

Naturally the questions arises, how this affects the derivation of the Hessian matrix of the Regge action. As it turns out, it hardly does. Let us prove it now:

The starting point of the derivation remains the same, namely we consider a simplicial complex made up of $d+2$ points embedded in $\mathbb{R}^d$ with non--degenerate $d$--simplices. Hence the Cayley--Menger determinant $\det C$ is vanishing and the Cayley--Menger matrix $C$ has exactly one null eigenvector. The embeddebility into $\mathbb{R}^d$ and also the vanishing of the Cayley--Menger determinant is due to the vanishing of the deficit angles $\omega_h$ (modulo $2\pi$), such that we require that the deficit angles do not change under deviations of the edge lengths, i.e. $\delta \omega_h \equiv 0$.

To put it differently, the condition $\det C=0$, where $C$ only has one null eigenvector, ensures that the given lengths form geometric $d$--dim. flat configurations. Thus, independent of the details of the definition of the deficit angles, we have:
\begin{equation}
 \delta \omega_{h}\, \neq \, 0\implies \delta\det C \, \neq \, 0\ .
\end{equation}
As a consequence, the variations of the deficit angles $\omega_h$ and $\det C$ are related. This shows that under the condition $\det C=0$, we conclude for the 4D deficit angles $\omega_{ijk}$:
\begin{equation}
 \delta \omega_{ijk}\,= \,B_{ijk} \, \delta\det C,\quad \frac{\partial\omega_{ijk}}{\partial l_{mn}}\,= \, B_{ijk} \, \frac{\partial\det C}{\partial l_{mn}} \quad ,
\end{equation}
where $B_{ijk}$ denote some implicit functions. It turns out that the bulk part of the Hessian satisfies \eqref{eq:4d-second}
\begin{equation}
 \frac{\partial^2 S_R}{\partial l_{op} \partial l_{mn}} = - \sum_h \frac{\partial A_h}{\partial l_{op}} \frac{\partial \omega_h}{\partial l_{mn}} + \text{bdry terms} = E_{op}\frac{\partial\det C}{\partial l_{mn}} + \text{bdry terms} \quad ,
\end{equation}
where $E_{op}$ again are some implicit functions. Since the Hessian matrix is symmetric, we can determine it up to an overall factor $F$:
\begin{equation}
 \delta^2 S_R\,=\, F\; \delta \det C\otimes \delta \det C+ \text{bdry terms} \quad .
\end{equation}
Since we only consider variations on the surface defined by $\det C = 0$, $\delta \det C$ is straightforward to determine:
\begin{equation}
 \frac{\partial\det C}{\partial l_{mn}}= \left. \text{Tr}\right|_{\det C = 0} \left(\text{adj}(C) \frac{\partial C}{\partial l_{mn}}\right)=\left( \vec{v}_c \Big|\frac{\partial C}{\partial l_{mn}} \vec{v}_c\right)=
 2^{-2} (4!)^{-2} \, l_{mn} (-1)^{s_m}V_{\bar{m}} (-1)^{s_n}V_{\bar{n}} \quad .
\end{equation}
Finally,
\begin{equation} \label{eq:detC-calc}
 \delta^2 S_R\,= \, F \; \frac{(-1)^{s_i + s_j + s_m + s_n}}{48^2} \, l_{ij} l_{mn} \,  V_{\bar{i}} V_{\bar{j}}  V_{\bar{m}}  V_{\bar{n}} \, \delta l_{ij}\otimes \delta l_{mn}
 + \text{ bdry terms} \quad .
\end{equation}
In order to identify $F$, let us repeat the reasoning from
\cite{steinhaus11}. Let us denote by $\{ijkmno\}$ a permutation of the vertices $\{012345\}$. It is straightforward to compute $\frac{\partial \omega_{ijk}}{\partial l_{mn}}$, since it only depends on the dihedral angle in the 4--simplex $(ijkmn)$. Using a formula from \cite{dittrich07} for $\frac{\partial \omega_{ijk}}{\partial l_{mn}}$, we get:
\begin{equation}
 \frac{\partial\omega_{ijk}}{\partial l_{mn}}=(-1)^{s_o}\frac{l_{mn}A_{ijk}}{12 V_{\bar{o}}}\Rightarrow
 \delta\omega_{ijk}=(-1)^{\sum_l s_l}\frac{(-1)^{s_i}V_{\bar{i}} \, (-1)^{s_j}V_{\bar{j}} \, (-1)^{s_k}V_{\bar{k}} \, A_{ijk}}{12 \prod_l V_{\bar{l}}}\
 \underbrace{(-1)^{s_m}V_{\bar{m}} \, (-1)^{s_n}V_{\bar{n}} \, l_{mn} \, \delta l_{mn}}_{ 48^2 \delta \det C} \quad .
\end{equation}
Thus in any case we have (for edge lengths not on the boundary)
\begin{equation}
 \frac{\partial^2 S_R}{\partial l_{mn}\partial l_{ij}}= - \sum \frac{\partial A_{ijk}}{\partial l_{ij}} \frac{\partial \omega_{ijk}}{\partial l_{mn}}
 =- \frac{\left(\sum_k (l_{ik}^2+l_{jk}^2-l_{ij}^2) (-1)^{s_k}V_{\bar{k}}\right)}{96 \prod_l V_{\bar{l}}}
 \underbrace{(-1)^{s_i}V_{\bar{i}}(-1)^{s_j}V_{\bar{j}}l_{ij}}_{48^2 \frac{\partial \det C}{\partial l_{ij}}}
 \underbrace{(-1)^{s_m}V_{\bar{m}}(-1)^{s_n}V_{\bar{n}}l_{mn}}_{48^2 \frac{\partial \det C}{\partial l_{mn}}} \quad ,
\end{equation}
where we used the fact that
\begin{equation}
 \frac{\partial A_{ijk}}{\partial l_{ij}}=\frac{l_{ij}}{8 \, A_{ijk}}(l_{ik}^2+l_{jk}^2-l_{ij}^2) \quad .
\end{equation}
We can thus identify $F=-24\frac{D}{\prod_{l} V_{\bar{l}}}$.

In this section, we have thoroughly discussed the changes of orientation that occur, if one moves the subdividing vertex outside the coarser 4--simplex. This vertex can be moved outside in different ways, which can be summarized by stating through which subsimplex it has `left' the coarser 4--simplex. Then all 4--simplices sharing this subsimplex change their relative orientation, which directly translates into a change of sign in the definition of the deficit angles for all dihedral angles stemming from these 4--simplices. Yet these intricacies of the deficit angles do not interfere with the derivations of \cite{steinhaus11} as long as the configuration is embeddable in $\mathbb{R}^4$, which is equivalent to stating that the Cayley--Menger determinant $\det C=0$ or the deficit angles $\omega_h = 0$ modulo $2 \pi$.  

We have the same factor $D$ appearing in the Hessian of the action, which is vanishing if the subdividing vertex is moved onto the 3--sphere defined by the five vertices of the coarser 4--simplex. In this case the fluctuation matrix $ \delta^2 S_R$ becomes singular\footnote{In case of the 5--1 move the Hessian matrix is singular, since it possesses four null eigenvectors corresponding to the vertex translation symmetry of the subdividing vertex. If additionally the factor $D$ vanishes, the whole Hessian matrix vanishes.}. Interestingly there is a recent conjecture, that not including the sum over orientations (and thus avoiding this situation), would avoid divergences in spin foams \cite{antispacetime}. The findings here support this conjecture in a further way: not only may the sum over orientations lead to non--compact (potential) gauge orbits. Additionally there is a submanifold of configurations, on which the fluctuation matrix becomes singular, which only appears if the subdividing vertex is moved outside the coarser simplex (and if one adds up the actions for the different simplices with their correct orientations).

\subsection{A $D$ factor absorbing measure for the 5--1 move}

Given the concise version of $D$ in affine coordinates, see \eqref{eq:D-affine}, one may ask whether one can construct a measure factor that  absorbs the $D$ factor under the 5--1 move. 
That is one needs to know which five simplices are coarse grained to one simplex $\sigma_{\bar{0}}$ as the factor $D$ refers to this simplex $\sigma_{\bar{0}}$. The measure one has to choose is then
 %
 %
 %
 %
 %
\begin{equation}
\mu(\{l\})_{5-1} = 2 b \sqrt{V_{\bar{0}}} \frac{\prod_e \frac{1}{\sqrt{192 \pi}} \, l_e}{\prod_i \sqrt{V_{\bar{i}}}} \quad \text{, with } b^2 = \sum_{0<i<j} \alpha_i \alpha_j l_{ij}^2 \quad ,
\end{equation}
where the $\alpha_i$ can be expressed in terms of the lengths $l_{i0}$ and the lengths of the simplex $\sigma_{\bar{0}}$.  Again, `non--locality' arises because it is impossible to rewrite this expression into factors that would only refer to the initial five 4--simplices (or other subsimplices).  Instead we have one factor referring to a complex of five simplices.


Here the modifications to the measure due to the non--local factor are specifically made to absorb these contributions, which still requires to identify (and specify) the complexes to which a 5--1 move can (and will) be applied. Thus subdividing one 4--simplex arbitrary many times by 1--5 moves and integrating out again by 1--5 moves with the measure above we will finally end up with an amplitude for the one final simplex, which of course will be `local', i.e. only refer to the geometry of this simplex.

One can nevertheless ask whether there exists a local measure invariant under a 5--1 move, restricted to special configurations. An example of such a restriction is to consider only barycentric subdivisions. The ansatz of a local measure (factorizing over (sub)simplices) leads to a functional equation that has to be solved, yet whether a solution exists is an open question. In appendix \ref{app:bary}, we present a measure
\begin{equation} 
\mu(\{l\}) =  \frac{\prod_e \frac{2^{\frac{1}{10}} 5^{\frac{1}{8}}}{\sqrt{192 \pi}} \, l^{6/5}_e}{\prod_\Delta V^{3/8}_\Delta}\quad 
\end{equation}
that is preserved under a single barycentric 5--1 move, if the coarse simplex is equilateral. This local factor should be taken with a grain of salt: Although it gives invariance for a subdivided equilateral simplex, it will not be invariant under repeated subdivisions, as equilateral simplices do not stay equilateral under subdivisions.

\section{Discussion} \label{sec:discussion}

In this note we have revisted the work in \cite{steinhaus11}, in which it has been examined whether one can construct a triangulation invariant path integral measure for (linearized) Regge calculus. While this is possible for the 3--dimensional topological theory, the 4--dimensional case is complicated by the appearance of an overall non--trivial factor $D$, see \eqref{eq:factor-D}, which cannot be factorized over  (sub)simplices of the triangulation. As we have shown in this paper, this factor has a peculiar geometric interpretation, which is the key ingredient to uncover its non--factorising and, therefore, non--local nature.

This result was derived for the linearized theory. It however also excludes a local measure both invariant under $5-1$ moves and gauge invariant in the sense described in the following for the full theory. The classical equation of motions for this move, which determine the solutions to be flat, display diffeomorphism symmetry as the position of the subdividing vertex can be anywhere (if it is outside the coarser simplex one needs to take the change of (relative) orientation into account). One thus needs to gauge fix (as in the linearized theory). If we assume that a local gauge invariant measure exists that leads to invariance under $5-1$ moves we could use this to define a local measure for the linearized theory, which however does not exist.

We suggested a non--local measure that would absorb the non--local $D$ factor under $5-1$ moves. Alternatively one can devise measure factors that are local and lead to an approximate invariance near very symmetric configurations. Such measure factors could be taken as a hint for choosing the measure for spin foams, see for instance \cite{sebastianwojtekBC} for a first geometric interpretation of the measure factor for a 4D model.

We additionally found that allowing for a sum over orientations (in our case having simplices of different orientations as background), can lead to more singular Hessians\footnote{In fact, the Hessian vanishes entirely if $D=0$.} for the linearized Regge action. This resonates with the conjecture in \cite{antispacetime}, that not summing over orientations might avoid divergences in spin foams, see however the discussion in \cite{time-evo}, pointing out the significance of summing over orientations for refining boundary states in gravity.

One could have hoped to find a measure that makes the path integral for the linearized theory invariant under $5-1$ moves (and also $4-2$ moves) and is local \cite{freidelprivate}. After all this kind of invariance holds for the Regge action and it was the initial motivation for the work \cite{steinhaus11}. This work indeed turned out to reproduce successfully the measure factor found in the Ponzano--Regge asymptotics \cite{pr-asymptotics}, the (topological) spin foam model for 3D gravity. We see that a theory with propagating degrees of freedom can have quite different properties in this respect even for a sector that leads to only flat solutions as is the case for the $5-1$ and $4-2$ move. Such a flat sector was also discussed in \cite{dittrichryan08,bonzomdittrich13} from a canonical viewpoint where the flatness indeed allows for anomaly free discretizations of the constraint algebra for a special class of boundaries. That is the Hamiltonian and diffeomorphism constraints can be defined, are local and are first class, which so far is not possible to achieve for the general 4D (discrete) case. The results in the work at hand question the possibility to find an anomaly free quantization for this flat (in a sense topological, as the boundaries do not allow for propagating degrees of freedom) sector with only local constraints.

We now have to expect that it is not possible to find a local  invariant measure (i.e. a one--loop effective action), even if one just wants to achieve invariance under $5-1$ moves. 
In fact  the concept of `invariance' under local moves involving  non--local amplitudes is rather hard to define:  the  amplitude for the initial complex in the 5--1 move is non--local, as it can be non--factorizing over the 5 simplices, however the final amplitude only refers to the final, coarser simplex. We therefore suggested a measure that would absorb the non--local part, which results from the Hessian of the action (where 'non--local part' is not without ambiguities).  The difficulties of formulating invariance conditions for non--local amplitudes arise also because one is keeping simplices as fundamental building blocks and the principle that these simplices are `glued'  together by integrating over boundary variables. This can be taken as a hint that an alternative formulation, as presented in  \cite{time-evo}, is worthwhile:  This formulation replaces simplices by building blocks with arbitrarily complicated boundaries and focuses on the amplitudes associated to these boundaries.  The requirement of triangulation invariance (which has become empty as it refers to the bulk, whereas in this formulation everything is defined via boundary geometries) is replaced by the condition of cylindrical consistency for the amplitudes associated to these boundaries \cite{dittrich-cyl}.   Such a formulation is much better suited for situations where non--local amplitudes arise and moreover does not refer to bulk triangulations at all. 
 
  On the level of the (classical) action, this non--locality has already to be expected if one wants to achieve full triangulation invariance \cite{he}, i.e. under all Pachner moves, as is the case with other perfect actions \cite{perfect-action}.

Our arguments relied on the Regge action, using length variables. An interesting question is whether these findings would change if one would use other variables, e.g. involving angles and / or areas \cite{barrettfirstorder,area-regge}. A first order formulation with length and (4D dihedral)  angles as in \cite{barrettfirstorder} would also lead to a non--local measure, as the angles can be integrated out in each simplex locally, and a local measure in first order variables would lead to a local measure in length variables. Area--Regge calculus itself has non--local constraints \cite{williamsarea}, thus one would even expect a non--local measure (for the associated Lagrange multipliers) for this reason. In contrast, area--angle Regge calculus \cite{area-regge} has local constraints, which become non--local (reducing to the area constraints) after integrating out the angles. Here the question of non--locality might depend on whether to impose the gluing (or shape matching) constraints, which are part of the constraints reducing are--angle Regge calculus to length Regge calculus \cite{area-regge}. The status of these constraints in spin foams (or loop quantum gravity) is debated, see \cite{jimmy2} for a discussion.
 Let us also mention \cite{baratin4D}, which finds a local measure for a 4D model with a Regge like (first order) action. This model is however topological, and hence it is not surprising that one can find a local measure factor in this case. It also shows that the issue of imposing gluing constraints, that are essential in regaining standard Regge calculus with propagating degrees of freedom, might be crucial.

The encountered non--localities are characteristic of interacting theories and unavoidable, but might be more effectively handled if other degrees of freedom are used. 
One possibility are dynamically determined recombinations of the initial degrees of freedom to obtain new effective degrees of freedom, which can be non--local in terms of the initial triangulation, but may interact only locally among each other. This can also be interpreted as combining the initial basic building blocks into new effective ones, which may require transformations that are more general than Pachner moves. These transformations can be for example be given by dynamical embedding maps \cite{dittrich-cyl,time-evo}, which relate the Hilbert spaces of finer and coarser boundary data. In fact the idea is to use these embedding maps to define the physical (continuum) Hilbert space via an inductive limit and requiring cylindrical consistency. Interestingly, these ideas naturally translate to ideas and real--space renormalization schemes in condensed matter physics, such as tensor network renormalization \cite{tnw}, which recently have been applied to analogue spin foam models \cite{tnw-qg}.

\begin{acknowledgments}
W.K. acknowledges partial support by the grant ``Maestro'' of  Polish Narodowe Centrum Nauki nr 2011/02/A/ST2/00300 and the grant of Polish Narodowe Centrum Nauki number 501/11-02-00/66-4162. S.St. gratefully acknowledges support by the DAAD
(German Academic Exchange Service) and would like to thank Perimeter Institute for an Isaac Newton
Chair Graduate Research Scholarship. This research was supported in part by Perimeter Institute for Theoretical Physics. Research at Perimeter Institute is supported by the Government of Canada through Industry Canada and by the Province of Ontario through the Ministry of Research and Innovation.
\end{acknowledgments}

\begin{appendix}

\section{Special cases of $D$}\label{sec:appendix}

\subsection{Placing the subdividing vertex far outside} \label{app:farout}

\begin{figure}[h!]
\begin{center}
\includegraphics[scale=1]{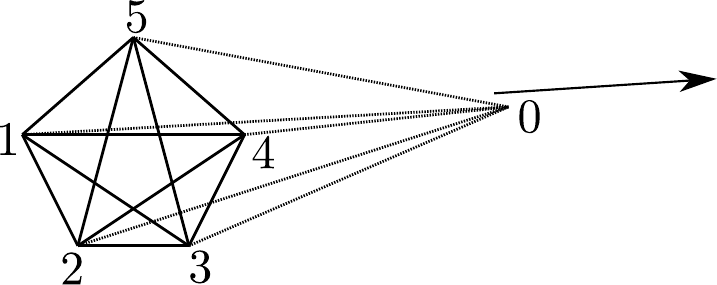}
\caption{\label{fig:far_out}
A schematic picture of the limit, in which the vertex $0$ is moved far away from the initial $4$--simplex.}
\end{center}
\end{figure}

In special situations it is possible to derive simpler expressions for the factor $D$. Assume for example that the inner vertex $0$ in the $5$--$1$ move is moved very far outside the original simplex as sketched in fig. \ref{fig:far_out}. These configurations, called `spikes', are important in determining the divergence behaviour of spin foam models \cite{perini,aldo,bonzomdittrich} and hence it is interesting to look for a simpler form of the factor $D$ in this limit.  
If sufficiently far out, we can approximate the new edge lengths $l_{0i} \approx l$. Inserting this into \eqref{eq:main}, we get:
\begin{equation}
 D^2 \approx \left(\frac{1}{48}\right)^2 l^4
\det\left(
\begin{matrix}
0 & 1 & \dots & 1 \\
1 & 0 & \dots & (l_{1(d+1)})^2 \\
\vdots & \vdots & \dots & \vdots \\
1 & (l_{1(d+1)})^2 & \dots & 0
\end{matrix}
\right)= 4 l^4 V_{\bar{0}}^2 \quad .
\end{equation}

\subsection{Circumcentric subdivision} \label{app:circum}

The same argument works out exactly in a circumcentric subdivision of the initial 4--simplex. In this subdivision, the subdividing vertex has the same distance to all vertices of the initial 4--simplex, i.e. it is placed in the center of circumscribing sphere of the initial simplex. Depending on the shape of the initial simplex, this can mean that the new vertex is inside or outside the simplex, yet by definition, it cannot be on the circumscribing sphere. In the following we denote the radius of the sphere as $r$, thus $l_{0i}=r$:
\begin{equation}
 D^2= \left(\frac{1}{48}\right)^2 \det\left(
 \begin{array}{ccccc}
  0 &r^2 &r^2&\cdots & r^2\\
  r^2& 0 & l_{12}^2 &\cdots & l_{15}^2\\
  r^2 & l_{12}^2 & 0 & \cdots & l_{25}^2\\
  \vdots & \vdots & \vdots &\ddots &\vdots\\
  r^2 & l_{15}^2 & \cdots &\cdots &0
 \end{array}\right)
 = 4 r^4 V_{\bar{0}}^2
\end{equation}
In both these cases the non--locality of $D$ is rather hidden in the simplicity of the considered geometry.

\section{Measure for barycentric subdivision} \label{app:bary}
 
Here, we will consider a barycentric subdivision, which in affine coordinates is given by choosing all $\alpha_i$ to be equal. In 4D, $\alpha_i=\frac{1}{5}$ and many equations given above are simplified:
\begin{equation}
b^2 = \frac{1}{25} \sum_{0<i<j} l_{ij}^2 \quad , \quad
l_{0k}^2 = \frac{1}{25} \left( 4 \sum_{i\neq k} l_{ik}^2 - \sum_{m,n \neq k} l_{mn}^2 \right) \quad .
\end{equation}
 
In case one considers an equilateral initial 4--simplex, i.e. all edge lengths $l_{ij} = l,\; \forall i,j \neq 0$, the factor $D$ simplifies even further. In fact, for this 4--simplex, the circumcentric subdivision coincides with the barycentric subdivision, $b^2 = r^2$. Then the subdividing edges also have the same length $l'$, with $(l')^2 := b^2 =\frac{2}{5} l^2$ and each of the five 4--simplices $\bar{i}, \; i \neq 0$, has the same volume $V'$, namely $V' := V_{\bar{i}} = \frac{1}{5} V_{\bar{0}} \; \forall i \neq 0$. We consider again $D_{01}$:
\begin{equation}
D_{01} = \sum_{k = 2}^5 \underbrace{(-1)^{s_k}}_{=1} (l_{0k}^2 + l_{1k}^2 - l_{01}^2) V_{\bar{k}} = \frac{4}{5} l^2 V_{\bar{0}} = 10 \, (l')^2 \, V' \quad .
\end{equation}
Thus we can construct a local measure, which is (approximately) invariant for integrating out subdivided (approximately) equilateral simplices. 

Consider the following measure\footnote{Here we choose to assign numerical factors to the edges as in \cite{steinhaus11}. Other choices are possible as well.}
\begin{equation} 
\mu(\{l\}) =  \frac{\prod_e \frac{2^{\frac{1}{10}} 5^{\frac{1}{8}}}{\sqrt{192 \pi}} \, l^{6/5}_e}{\prod_\Delta V^{3/8}_\Delta}\quad 
\end{equation} 
for the initial complex of five simplices in the (equilateral, barycentric) $5-1$ move. From this $5-1$ move integration we obtain the following factors from the (square root of the inverse of the determinant of the) Hessian and the gauge fixing procedure described in \cite{steinhaus11}, by which the initial measure factor is multiplied:
\begin{equation} 
F_{5-1} = (192 \pi)^{\frac{5}{2}} \,  \frac{1}{\sqrt{D}} \sqrt{  \frac{V_{\bar{1}} \cdots V_{\bar{5}}}{V_{\bar{0}}}}  \frac{1}{l_{01}\cdots l_{05}} \quad .
\end{equation} 
For the equilateral barycentric $5-1$ move we thus obtain
\begin{equation}
\mu_{\text{initial}} F_{5-1} \,=\, \frac{2^{\frac{3}{2}} 5^{\frac{15}{8}}}{(192 \pi)^{\frac{15}{2}}} \; (192 \pi)^\frac{5}{2} \; \frac{1}{5^\frac{5}{8}}  \frac{1}{2^{\frac{1}{2}} } \, \frac{ l^{\frac{6}{5} \cdot 10}}{ V_{\bar{0}}^{\frac{3}{8}} } \,\,=\,\, \mu_{\text{final}}
\end{equation} 
and hence invariance for this highly symmetric configuration. Note that this procedure is not without ambiguities, apart from the question of how to distribute numerical coefficients, we could have also exchanged length variables for volumina and vice versa using the relation $V_{\bar{0}}=  \sqrt{5}\,(2^3\cdot 3)^{-1}  l^4$ for the equilateral 4--simplex.

\section{Radius of circumscribing sphere} \label{app:c-radius}

The radius $r$ of the $(d-1)$--sphere $S$ circumscribing the $d$--simplex $\Delta$ can be computed from the Cayley--Menger matrix $C$ of $\Delta$ \cite{berger} (see also \cite{dandrea}):
\begin{equation}
r(S) = -\frac{1}{2} \frac{\left|C_0 (\Delta) \right|}{\det C(\Delta)} \quad ,
\end{equation} 
where the numerator $\left|C_0(\Delta) \right|$ is the determinant of the `$(0,0)$' minor of $C(\Delta)$, see also \eqref{eq:D-Cayley}, and the denominator is simply the Caley--Menger determinant of $\Delta$. See section \ref{sec:orientation} for the notation.
\end{appendix}


\begin{thebibliography}{99}\small
\parskip -1pt

\bibitem{spinfoams}
  A.~Perez,
  ``The Spin Foam Approach to Quantum Gravity,''
  Living Rev.\ Rel.\  {\bf 16} (2013) 3
  [arXiv:1205.2019 [gr-qc]].
  
  C.~Rovelli,
  ``Zakopane lectures on loop gravity,''
  PoS QGQGS {\bf 2011} (2011) 003
  [arXiv:1102.3660 [gr-qc]].

\bibitem{gft}
  D.~Oriti,
  ``The Group field theory approach to quantum gravity,''
  In *Oriti, D. (ed.): Approaches to quantum gravity* 310-331
  [gr-qc/0607032].
  
  R.~Gurau and J.~P.~Ryan,
  ``Colored Tensor Models - a review,''
  SIGMA {\bf 8} (2012) 020
  [arXiv:1109.4812 [hep-th]].
  
\bibitem{cdt}
J.~Ambjorn, A.~Goerlich, J.~Jurkiewicz and R.~Loll,
  arXiv:1302.2173 [hep-th].
  
\bibitem{regge}
T.~Regge,
 ``General relativity without coordinates,'' Nuovo Cim. {\bf 19} (1961) 558;
 
R. M. Williams, 
``Recent progress in Regge calculus,'' Nucl. Phys. Proc. Suppl. {\bf 57} (1997) 73 [arXiv:gr-qc/9702006].

M.~Rocek and R.~M.~Williams,
  ``Quantum Regge Calculus,''
  Phys.\ Lett.\ B {\bf 104} (1981) 31.
  
  M.~Rocek and R.~M.~Williams,
  ``The Quantization of Regge Calculus,''
  Z.\ Phys.\ C {\bf 21} (1984) 371.

%

\bibitem{schrader}
J. Cheeger, W. M\"uller and R. Schrader,
``On the curvature of piecewise flat spaces,''
Comm. Math. Phys. {\bf 92} (1984) 405-454.

\bibitem{hamber}
H.~W.~Hamber and R.~M.~Williams,
  ``On the measure in simplicial gravity,''
  Phys.\ Rev.\ D {\bf 59} (1999) 064014
  [hep-th/9708019].

\bibitem{hamber2}
H.~W.~Hamber and R.~M.~Williams,
  ``Gauge invariance in simplicial gravity,''
  Nucl.\ Phys.\ B {\bf 487} (1997) 345
  [hep-th/9607153].

\bibitem{menotti}
P.~Menotti and P.~P.~Peirano,
  ``Diffeomorphism invariant measure for finite dimensional geometries,''
  Nucl.\ Phys.\ B {\bf 488} (1997) 719
  [hep-th/9607071].

\bibitem{regge-lund}
F. Lund and T. Regge, Princeton preprint (1974), unpublished.

\bibitem{bojowald}
 M.~Bojowald and A.~Perez,
  ``Spin foam quantization and anomalies,''
  Gen.\ Rel.\ Grav.\  {\bf 42} (2010) 877
  [gr-qc/0303026].
  
\bibitem{kkl}
W.~Kaminski, M.~Kisielowski and J.~Lewandowski,
  ``Spin-Foams for All Loop Quantum Gravity,''
  Class.\ Quant.\ Grav.\  {\bf 27} (2010) 095006
   [Erratum-ibid.\  {\bf 29} (2012) 049502]
  [arXiv:0909.0939 [gr-qc]].

\bibitem{diffeo-review}
 B.~Dittrich,
  ``Diffeomorphism symmetry in quantum gravity models,''
  Adv.\ Sci.\ Lett.\  {\bf 2} 151
  [arXiv:0810.3594 [gr-qc]].

\bibitem{broken-symmetry}
  B.~Bahr and B.~Dittrich,
  ``(Broken) Gauge Symmetries and Constraints in Regge Calculus,''
  Class.\ Quant.\ Grav.\  {\bf 26} (2009) 225011
  [arXiv:0905.1670 [gr-qc]].

  B.~Bahr and B.~Dittrich,
  ``Breaking and restoring of diffeomorphism symmetry in discrete gravity,''
  arXiv:0909.5688 [gr-qc].

\bibitem{louapre}
L.~Freidel and D.~Louapre,
  ``Diffeomorphisms and spin foam models,''
  Nucl.\ Phys.\ B {\bf 662} (2003) 279
  [gr-qc/0212001].

\bibitem{bahr}
 B.~Bahr,
  ``On knottings in the physical Hilbert space of LQG as given by the EPRL model,''
  Class.\ Quant.\ Grav.\  {\bf 28} (2011) 045002
  [arXiv:1006.0700 [gr-qc]].
  
\bibitem{warsaw}
B.~Bahr, F.~Hellmann, W.~Kaminski, M.~Kisielowski and J.~Lewandowski,
  ``Operator Spin Foam Models,''
  Class.\ Quant.\ Grav.\  {\bf 28} (2011) 105003
  [arXiv:1010.4787 [gr-qc]].

\bibitem{bonzomdittrich}
V.~Bonzom and B.~Dittrich,
  ``Bubble divergences and gauge symmetries in spin foams,''
  Phys.\ Rev.\ D {\bf 88} (2013) 124021
  [arXiv:1304.6632 [gr-qc]].
  
\bibitem{perini}
C.~Perini, C.~Rovelli and S.~Speziale,
  ``Self-energy and vertex radiative corrections in LQG,''
  Phys.\ Lett.\ B {\bf 682} (2009) 78
  [arXiv:0810.1714 [gr-qc]].

\bibitem{aldo}
  A.~Riello,
  ``Self-Energy of the Lorentzian EPRL-FK Spin Foam Model of Quantum Gravity,''
  Phys.\ Rev.\ D {\bf 88} (2013) 024011
  [arXiv:1302.1781 [gr-qc]].

\bibitem{deWitt}
B.~S.~DeWitt,
  ``Quantum Theory of Gravity. 1. The Canonical Theory,''
  Phys.\ Rev.\  {\bf 160} (1967) 1113.

\bibitem{steinhaus11}
B.~Dittrich and S.~Steinhaus,
  ``Path integral measure and triangulation independence in discrete gravity,''
  Phys.\ Rev.\ D {\bf 85} (2012) 044032
  [arXiv:1110.6866 [gr-qc]].

\bibitem{perfect-action}
  P.~Hasenfratz and F.~Niedermayer,
  ``Perfect lattice action for asymptotically free theories,''
  Nucl.\ Phys.\ B {\bf 414} (1994) 785
  [hep-lat/9308004].
  
  P.~Hasenfratz,
  ``Prospects for perfect actions,''
  Nucl.\ Phys.\ Proc.\ Suppl.\  {\bf 63} (1998) 53
  [hep-lat/9709110].

\bibitem{Regge-cosmo}  
  B.~Bahr and B.~Dittrich,
  ``Improved and Perfect Actions in Discrete Gravity,''
  Phys.\ Rev.\ D {\bf 80} (2009) 124030
  [arXiv:0907.4323 [gr-qc]].

\bibitem{marsden}
 J. Marsden, M. West, ``Discrete mechanics and variational integrators,'' Acta Numerica {\bf 10}
(2001) 357

\bibitem{anharmonic}
  B.~Bahr, B.~Dittrich and S.~Steinhaus,
  ``Perfect discretization of reparametrization invariant path integrals,''
  Phys.\ Rev.\ D {\bf 83} (2011) 105026
  [arXiv:1101.4775 [gr-qc]].

\bibitem{dittrich-cyl}
  B.~Dittrich,
  ``From the discrete to the continuous: Towards a cylindrically consistent dynamics,''
  New J.\ Phys.\  {\bf 14} (2012) 123004
  [arXiv:1205.6127 [gr-qc]].

\bibitem{time-evo}
  B.~Dittrich and S.~Steinhaus,
  ``Time evolution as refining, coarse graining and entangling,''
  arXiv:1311.7565 [gr-qc].
  
\bibitem{hoehn}
  B.~Dittrich and P.~A.~H\"ohn,
  ``Canonical simplicial gravity,''
  Class.\ Quant.\ Grav.\  {\bf 29} (2012) 115009
  [arXiv:1108.1974 [gr-qc]].
  
  B.~Dittrich and P.~A.~H\"ohn,
  ``Constraint analysis for variational discrete systems,''
  J.\  Math.\  Phys.\  {\bf 54}, (2013) 093505
  [arXiv:1303.4294 [math-ph]].
  
  P.~A.~H\"ohn,
  ``Quantization of systems with temporally varying discretization I: Evolving Hilbert spaces,''
  arXiv:1401.6062 [gr-qc].
  
  P.~A.~H\"ohn,
  ``Quantization of systems with temporally varying discretization II: Local evolution moves,''
  arXiv:1401.7731 [gr-qc].
  
\bibitem{pachner}
  U.~Pachner,
  ``Konstruktionsmethoden und das kombinatorische Hom\"oomorphieproblem f\"ur Triangulationen kompakter semilinearer Mannigfaltigkeiten,''
  Abh. Math. Sem. Univ. Hamburg {\bf 57} (1986) 69.
  
  U.~Pachner,
  ``P.L. Homeomorphic Manifolds are Equivalent by Elementary Shellings,''
  Europ. J. Combinatorics {\bf 12} (1991), 129-145.  
  
\bibitem{pr-asymptotics}
 G.~Ponzano and T.~Regge,
 {\it Semiclassical limit of Racah coefficients, in: Spectroscopy and group
 theoretical methods in physics}. 
 North Holland Publ. Co., Amsterdam, 1968.
 
 J.~W.~Barrett and I.~Naish-Guzman,
  ``The Ponzano-Regge model,''
  Class.\ Quant.\ Grav.\  {\bf 26} (2009) 155014
  [arXiv:0803.3319 [gr-qc]].
  
  R.~Gurau,
  ``The Ponzano-Regge asymptotic of the 6j symbol: An Elementary proof,''
  Annales Henri Poincare {\bf 9} (2008) 1413
  [arXiv:0808.3533 [math-ph]].
  
  M.~Dupuis and E.~R.~Livine,
  ``Pushing Further the Asymptotics of the 6j-symbol,''
  Phys.\ Rev.\ D {\bf 80} (2009) 024035
  [arXiv:0905.4188 [gr-qc]].
  
  M.~Dupuis and E.~R.~Livine,
  ``The 6j-symbol: Recursion, Correlations and Asymptotics,''
  Class.\ Quant.\ Grav.\  {\bf 27} (2010) 135003
  [arXiv:0910.2425 [gr-qc]].
  
  W.~Kaminski and S.~Steinhaus,
  ``Coherent states, 6j symbols and properties of the next to leading order asymptotic expansions,''
  J.\ Math.\ Phys.\  {\bf 54} (2013) 121703
  [arXiv:1307.5432 [math-ph]].
  
\bibitem{berger}
 M.~Berger,
 {\it Geometry I},
 Springer, Berlin, 1994  
  
\bibitem{dandrea}
C. D'Andrea and M. Sombra,
``The Cayley-Menger determinant is irreducible for $n \geq 3$,''
Siberian Mathematical Journal {\bf 46} (2005), pp 71-76
[math/0406359].  
  
\bibitem{dittrich07}
  B.~Dittrich, L.~Freidel and S.~Speziale,
  ``Linearized dynamics from the 4-simplex Regge action,''
  Phys.\ Rev.\ D {\bf 76} (2007) 104020
  [arXiv:0707.4513 [gr-qc]].
  
\bibitem{korepanov}
  I.~G.~Korepanov,
  ``Invariants of PL manifolds from metrized simplicial complexes: Three-dimensional case,''
  J.\ Nonlin.\ Math.\ Phys.\  {\bf 8} (2001) 196
  [math/0009225 [math-gt]].
  
  I.~G.~Korepanov,
  ``Multidimensional analogues of the geometric s -- t duality,''
  Theor.\ Math.\ Phys.\  {\bf 124} (2000) 999
   [Teor.\ Mat.\ Fiz.\  {\bf 124} (2000) 169].

\bibitem{baratin3d}
  A.~Baratin and L.~Freidel,
  ``Hidden Quantum Gravity in 3-D Feynman diagrams,''
  Class.\ Quant.\ Grav.\  {\bf 24} (2007) 1993
  [gr-qc/0604016].
 
\bibitem{sorkin}
  R.~Sorkin,
  ``The Electromagnetic field on a simplicial net,''
  J.\ Math.\ Phys.\  {\bf 16} (1975) 2432
   [Erratum-ibid.\  {\bf 19} (1978) 1800].
  
\bibitem{kokkendorff}
 S.~L.~Kokkendorff,
 ``Polar duality and the generalized law of sines,''
 Journal of Geometry {\bf 86} no. 1-2, (2007) 140--149.
  
\bibitem{antispacetime}
  M.~Christodoulou, M.~Langvik, A.~Riello, C.~Roken and C.~Rovelli,
  ``Divergences and Orientation in Spinfoams,''
  Class. Quantum Grav. 30 055009 2013
  [arXiv:1207.5156 [gr-qc]].

\bibitem{sebastianwojtekBC}
W.~Kamiński and S.~Steinhaus,
  ``The Barrett–Crane model: asymptotic measure factor,''
  Class.\ Quant.\ Grav.\  {\bf 31} (2014) 075014
  [arXiv:1310.2957 [gr-qc]].

\bibitem{freidelprivate}
L.~Freidel, private communication.

\bibitem{dittrichryan08}
B.~Dittrich and J.~P.~Ryan,
  ``Phase space descriptions for simplicial 4d geometries,''
  Class.\ Quant.\ Grav.\  {\bf 28} (2011) 065006
  [arXiv:0807.2806 [gr-qc]].
  
\bibitem{bonzomdittrich13}
V.~Bonzom and B.~Dittrich,
  ``Dirac’s discrete hypersurface deformation algebras,''
  Class.\ Quant.\ Grav.\  {\bf 30} (2013) 205013
  [arXiv:1304.5983 [gr-qc]].

\bibitem{he}
 B.~Bahr, B.~Dittrich and S.~He,
  ``Coarse graining free theories with gauge symmetries: the linearized case,''
  New J.\ Phys.\  {\bf 13} (2011) 045009
  [arXiv:1011.3667 [gr-qc]].
  
\bibitem{area-regge}
  B.~Dittrich and S.~Speziale,
  ``Area-angle variables for general relativity,''
  New J.\ Phys.\  {\bf 10} (2008) 083006
  [arXiv:0802.0864 [gr-qc]].
  
  B.~Bahr and B.~Dittrich,
  ``Regge calculus from a new angle,''
  New J.\ Phys.\  {\bf 12} (2010) 033010
  [arXiv:0907.4325 [gr-qc]].

\bibitem{barrettfirstorder}
 J.~W.~Barrett,
  ``First order Regge calculus,''
  Class.\ Quant.\ Grav.\  {\bf 11} (1994) 2723
  [hep-th/9404124].

\bibitem{williamsarea}
  J.~W.~Barrett, M.~Rocek and R.~M.~Williams,
  ``A note on area variables in Regge calculus,''
  Class.\ Quant.\ Grav.\  {\bf 16}, 1373 (1999)
  [arXiv:gr-qc/9710056].
    J.~Makela and R.~M.~Williams,
  ``Constraints on area variables in Regge calculus,''
  Class.\ Quant.\ Grav.\  {\bf 18}, L43 (2001)
  [arXiv:gr-qc/0011006].

\bibitem{jimmy2}
 B.~Dittrich and J.~P.~Ryan,
  ``Simplicity in simplicial phase space,''
  Phys.\ Rev.\ D {\bf 82} (2010) 064026
  [arXiv:1006.4295 [gr-qc]].
 B.~Dittrich and J.~P.~Ryan,
  ``On the role of the Barbero-Immirzi parameter in discrete quantum gravity,''
  arXiv:1209.4892 [gr-qc].

\bibitem{baratin4D}
 A.~Baratin and L.~Freidel,
  ``Hidden Quantum Gravity in 4-D Feynman diagrams: Emergence of spin foams,''
  Class.\ Quant.\ Grav.\  {\bf 24} (2007) 2027
  [hep-th/0611042].

\bibitem{tnw}
  M.~Levin and C.~P.~Nave,
  ``Tensor renormalization group approach to 2D classical lattice models,''
  Phys. Rev. Lett. {\bf 99}, (2007) 120601,
  [arXiv:cond-mat/0611687].
  
  Z.~Gu, X.~Wen,
  ``Tensor-Entanglement-Filtering Renormalization Approach and Symmetry Protected Topological Order,''
  Phys. Rev. B {\bf 80} (2009), 155131,
  [arXiv:0903.1069 [cond-mat.str-el]]
  
\bibitem{tnw-qg}
  B.~Dittrich, M.~Martín-Benito and E.~Schnetter,
  ``Coarse graining of spin net models: dynamics of intertwiners,''
  New J.\ Phys.\  {\bf 15} (2013) 103004
  [arXiv:1306.2987 [gr-qc]].
  
   B.~Dittrich, M.~Martin-Benito and S.~Steinhaus,
  ``Quantum group spin nets: refinement limit and relation to spin foams,''
  Phys.\ Rev.\ D {\bf 90} (2014) 024058
  [arXiv:1312.0905 [gr-qc]].

\end{thebibliography}
\end{document}